\documentclass[12pt,preprint]{aastex}

\shorttitle{Globular clusters in the outer halo of M31}
\shortauthors{Mackey et al.}

\begin{document}

\title{ACS photometry of newly-discovered globular clusters in the outer halo of M31\altaffilmark{1}}


\author{A.D. Mackey\altaffilmark{2}, A. Huxor\altaffilmark{3}, A.M.N. Ferguson\altaffilmark{2}, N.R. Tanvir\altaffilmark{4}, M. Irwin\altaffilmark{5}, R. Ibata\altaffilmark{6}, T. Bridges\altaffilmark{7}, R.A. Johnson\altaffilmark{8}, G. Lewis\altaffilmark{9}}


\altaffiltext{1}{Based on observations made with the NASA/ESA Hubble Space 
Telescope, obtained at the Space Telescope Science Institute, which is operated 
by the Association of Universities for Research in Astronomy, Inc., under NASA 
contract NAS 5-26555. These observations are associated with program 10394.}
\altaffiltext{2}{Institute for Astronomy, University of Edinburgh, Royal Observatory, 
Blackford Hill, Edinburgh, EH9 3HJ, UK}
\altaffiltext{3}{Department of Physics, University of Bristol, Tyndall Avenue, 
Bristol, BS8 1TL, UK}
\altaffiltext{4}{Department of Physics \& Astronomy, University of Leicester, 
Leicester, LE1 7RH, UK}
\altaffiltext{5}{Institute of Astronomy, University of Cambridge, Madingley Road, 
Cambridge, CB3 0HA, UK}
\altaffiltext{6}{Observatoire de Strasbourg, 11 rue de l'Universit\'{e}, F-67000 Strasbourg, France}
\altaffiltext{7}{Department of Physics, Queen's University, Kingston, ON, K7M 3N6, Canada}
\altaffiltext{8}{Oxford Astrophysics, Denys Wilkinson Building, Keble Road, Oxford, 
OX1 3RH, UK}
\altaffiltext{9}{Institute of Astronomy, School of Physics, A29, University of Sydney, NSW 2006, Australia}


\begin{abstract}
We report the first results from deep ACS imaging of ten classical
globular clusters in the far outer regions ($15 \la R_p \la 100$ kpc)
of M31. Eight of the clusters, including two of the most remote M31
globular clusters presently known, are described for the first
time. Our F606W, F814W colour-magnitude diagrams extend $\sim 3$
magnitudes below the horizontal branch and clearly demonstrate that
the majority of these objects are old ($\ga 10$ Gyr), metal-poor
clusters. Five have $[$Fe$/$H$] \sim -2.1$, while an additional four
have $-1.9 \la [$Fe$/$H$] \la -1.5$. The remaining object is more
metal-rich, with $[$Fe$/$H$] \sim -0.70$. Several clusters exhibit the
second parameter effect. Using aperture photometry, we estimate
integrated luminosities and structural parameters for all
clusters. Many, including all four clusters with projected radii
greater than $45$ kpc, are compact and very luminous, with $-8.9 \la
M_V \la -8.3$. These four outermost clusters are thus quite unlike their Milky
Way counterparts, which are typically diffuse, sub-luminous ($-6.0 \la
M_V \la -4.7$) and more metal-rich ($-1.8 \la [$Fe$/$H$] \la -1.3$).
\end{abstract}




\section{Introduction}
Globular clusters observed in M31 provide the closest example of a globular
cluster system belonging to a large external galaxy. Members of this system
allow us to trace the star formation history, chemical evolution, kinematics,
and mass distribution in different regions of M31, and are therefore
vital to our developing picture of its formation and evolution. However, very 
few clusters have been discovered at large distances from M31, meaning that up
until now we do not possess a good sample with which to probe its outer halo.
Recently, as part of a major survey of the M31 halo with the Isaac Newton Telescope 
\citep[see e.g.,][]{ferguson:02} and MegaCam on the Canada-France-Hawaii Telescope 
\citep[see e.g.,][]{martin:06}, a search for previously unknown globular clusters, 
at large projected radii ($R_p$) from the center of M31, has been conducted. This 
search has resulted in the discovery of a significant number of new clusters, with 
$R_p$ between $\sim 15-120$ kpc \citep{huxor:04,huxor:05,huxor:06,martin:06}.
A sample of $14$ of these objects has been the subject of deep follow-up imaging
using the Advanced Camera for Surveys (ACS) on board the Hubble Space Telescope
(HST). Preliminary results for four members of the population of luminous, extended 
clusters of \citet{huxor:05} have recently been described 
\citep[][hereafter Paper I]{mackey:06}. In this Letter, we present results 
for the remaining ten clusters, and compare, for the first time, the cluster 
population of the outer M31 halo to that of the outer Milky Way halo.

\begin{figure}
\begin{center}
\includegraphics[width=100mm,bb=82 95 590 645]{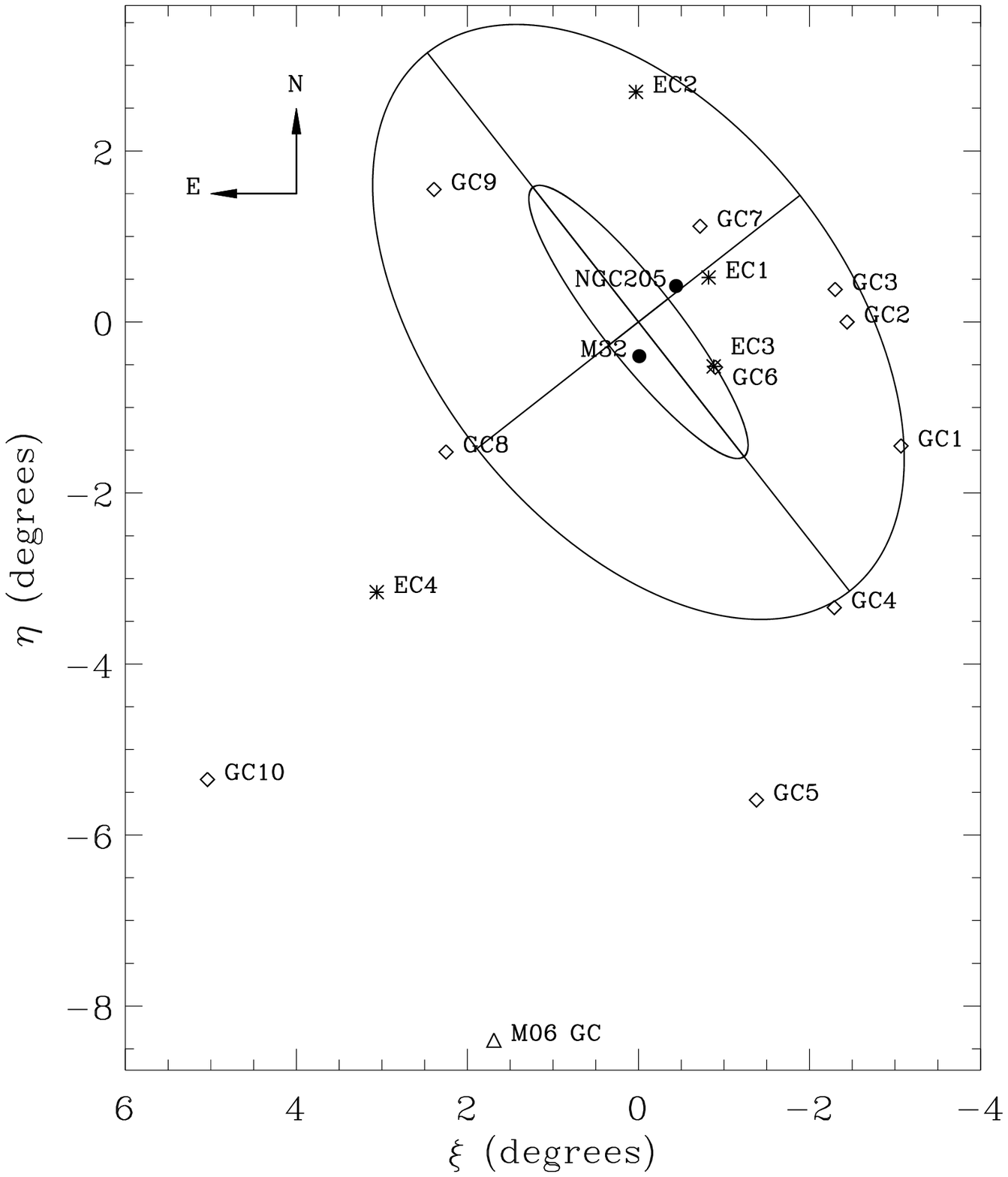} \\
\includegraphics[width=110mm]{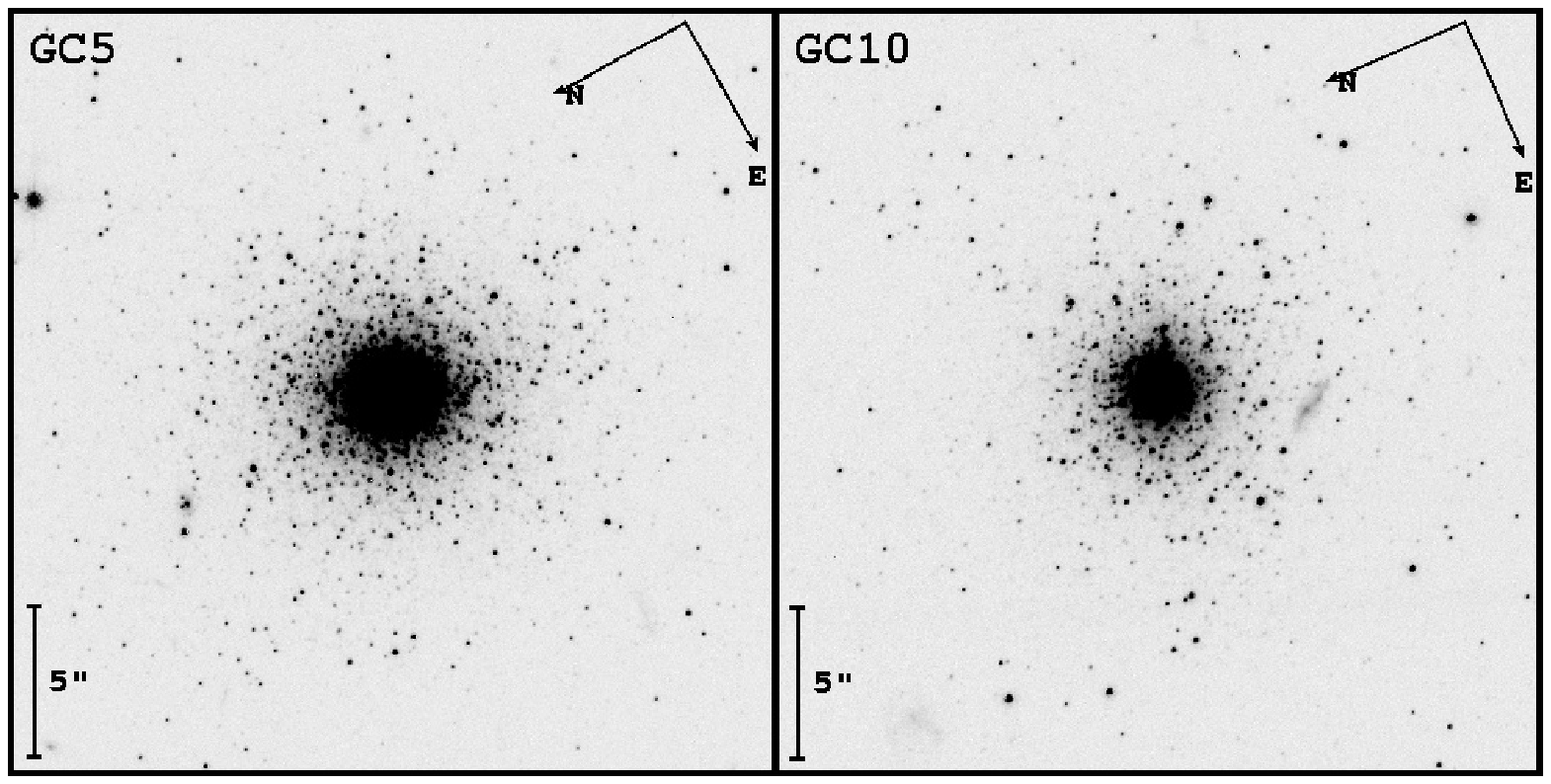}
\end{center}
\vspace{-3mm}
\caption{The upper panel shows the location of our present cluster sample (open
diamonds) in relation to M31. The four luminous, extended clusters from 
Paper I are marked with stars. Also displayed is the remote cluster of 
\citet{martin:06} (open triangle) and two M31 companion galaxies -- M32 and NGC 205 
(filled circles). The small ellipse delineates the visual extent of the M31 disk.
The lower panel shows drizzled ACS/WFC F606W images of the two outermost clusters
in our sample ($R_p \sim 78$ and $100$ kpc, respectively). Visually, they
are representative of the classical globular clusters considered in this paper. 
Each thumbnail has dimensions of $25\arcsec \times 25\arcsec$.\label{f:images}}
\end{figure}

This sample consists of classical (compact) globular clusters with projected
radii distributed in the range $15 \la R_p \la 100$ kpc. Only two of these
objects are previously recorded -- the remaining eight are new discoveries. Their 
positions are listed in Table \ref{t:results}, and presented schematically in
Fig. \ref{f:images}. Two of the newly-discovered clusters (GC5, GC10) lie at 
very large distances from M31: $R_p \sim 78$ and $100$ kpc, respectively. These,
along with the new cluster of \citet{martin:06}, which has $R_p \sim 116$ kpc,
are by far the most remote members of M31's globular cluster system found to date. 
Of the two previously reported clusters in our sample, one (GC4) was first described 
by \citet{huxor:04} but discovered independently by \citet{galleti:05}, who labelled 
it B514. This object has additional separate ACS imaging \citep{galleti:06}. The 
second was observed in the same field as one of our luminous, extended clusters, 
and appears in the catalogue of \citet{barmby:00} (cluster 298-021). 

\section{Observations and data reduction}
Our observations were obtained with the ACS Wide Field Channel (WFC) under 
HST program GO 10394 (P.I. Tanvir), at various intervals over the period
2005 May 27 -- 2005 September 17. Targets were imaged three times in
the F606W filter and four times in the F814W filter, with small dithers
between subsequent images. Typical total integration times were 1800s 
in F606W and 3000s in F814W. Each cluster was placed at the centre of 
either chip 1 or chip 2 on the WFC, to avoid the inter-chip gap. 
Drizzled F606W images of two representative clusters are displayed in 
Fig. \ref{f:images}. 

Details of our photometric reduction procedure are provided in Paper I.
Briefly, we used the {\sc dolphot} software \citep{dolphin:00}, 
specifically the ACS module, to conduct PSF-fitting photometry on all images. 
Output measurements are on the VEGAMAG scale of \citet{sirianni:05}. 
We used the quality information provided by {\sc dolphot} to clean the 
resulting detection lists, selecting only stellar detections,
with valid photometry on all input images, global sharpness parameter 
between $-0.3$ and $0.3$ in each filter, and crowding parameter less 
than $0.25$ in each filter. Field stars are present in all images; however,
their densities are typically sparse since the clusters are at such large
projected radii from M31. To eliminate field star contamination from our
cluster colour-magnitude diagrams (CMDs), we imposed limiting 
radii from the cluster centers. Because of the sparse fields, these radii were 
typically quite generous ($\sim 25\arcsec$). However, two clusters (GC6, GC7) are 
set against heavier background fields and hence required smaller limiting radii of 
$10\arcsec$. From the resulting CMDs (see below), it is evident that field
star contamination is not a significant issue -- in all cases the prinicpal
cluster sequences are very clearly visible. Therefore, a more sophisticated
statistical subtraction is not necessary at this stage.

\section{Analysis}
\begin{figure}
\epsscale{1.0}
\plotone{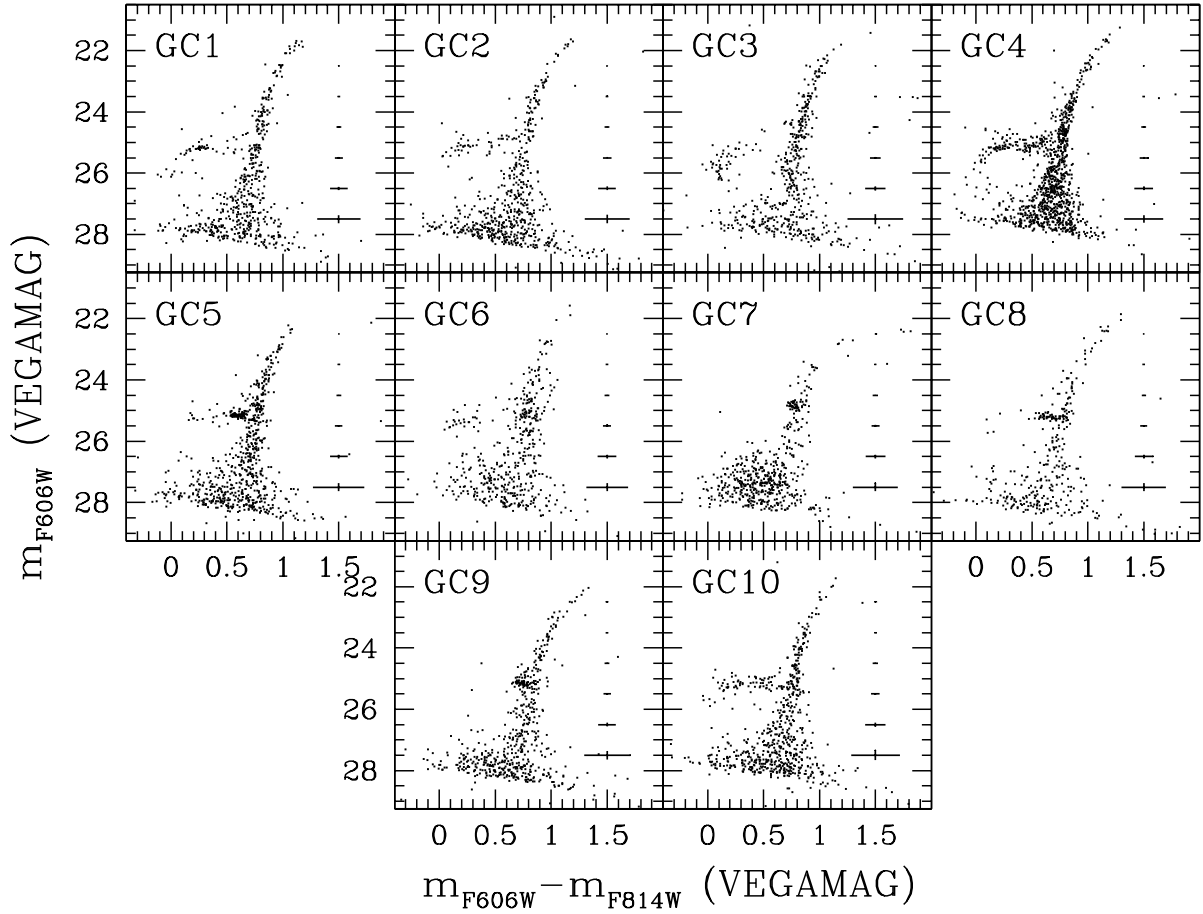}
\caption{CMDs for the ten classical M31 globular clusters. Photometry has been selected from the full ACS/WFC fields by imposing radial limits as described in the text. Typical photometric errors from {\sc dolphot} are indicated.\label{f:cmds}}
\end{figure}

Fig. \ref{f:cmds} shows CMDs for our ten clusters. The photometry
reaches $\sim 3$ mag below the level of the horizontal branch (HB), to a limiting
magnitude of $m_{{\rm F606W}} \sim 28$. All clusters exhibit narrow red-giant 
branches (RGBs) and clearly delineated HBs. Nine of the clusters possess rather
steep RGBs, indicating they are metal-poor objects. Many of these clusters
also feature HBs extending to the blue, including broadened regions with colours in
the range $0.1 \la m_{{\rm F606W}} - m_{{\rm F814W}} \la 0.5$ which are suggestive
of RR Lyrae stars imaged at random phase. Such clusters are old, with
ages $\ga 10$ Gyr. GC7 has a noticeably different CMD from those of the other clusters, 
with a sharply bending RGB and very stubby red HB, characteristics indicative of 
a more metal-rich object.

We obtained photometric metallicity estimates using a procedure we previously 
described and verified in Paper I. Because our sample covers a very wide spatial 
area and range in $R_p$, we could not assume a uniform distance and reddening. 
Instead, we determined the best-fitting combination of $[$Fe$/$H$]$, $(m-M)_0$, and 
$E(B-V)$ for a given cluster by registering the ACS/WFC F606W and F814W Galactic
globular cluster fiducials of \citet{brown:05} to the appropriate CMD using the 
F606W level of the HB and the colour of the RGB at the HB level. A fiducial of the 
correct $[$Fe$/$H$]$ closely traces the upper RGB on the CMD when registered in this 
way, while a fiducial of incorrect $[$Fe$/$H$]$ deviates on the upper RGB. The cluster 
metallicity is estimated by bounding the RGB with a more metal-poor and a more 
metal-rich fiducial. We assume {\it a priori} that the main body of M31 has 
$(m-M)_0 = 24.47 \pm 0.07$ \citep{mcconnachie:05}. For each cluster we registered 
fiducials using, incrementally, a range $\pm 0.5$ mag about this value (i.e., a 
system depth of $\sim 360$ kpc). At each $(m-M)_0$ we used Brown et al's 
transformations to solve for the $E(B-V)$ which aligned the fiducial and CMD HB 
levels. Applying this reddening, we determined the offset between the color of the
fiducial RGB at HB level and that of the cluster CMD. The best fitting combination
of $(m-M)_0$ and $E(B-V)$ for a given fiducial was that which minimized this
offset. To determine uncertainties in these values we randomly selected
ten thousand Gaussian deviates about the two HB levels and the RGB colour
at HB level (widths were typically $\pm 0.1$ mag and $\pm 0.01$ mag,
respectively) and calculated new $E(B-V)$ and $(m-M)_0$ for each.
We derived standard random errors from the resulting distributions.
We note that there are possibly additional systematic errors present for our 
measured $E(B-V)$ values, due to differences in age between the M31 clusters 
and the fiducial clusters, and the fact that the colour of the RGB at the HB level 
is mildly age-sensitive. For the majority of clusters (with the exceptions discussed 
below), these errors would amount at most to $\sim 0.02$ mag more than quoted in Table 
\ref{t:results}, corresponding to the interval $10-15$ Gyr.

\begin{figure}
\epsscale{1.0}
\plotone{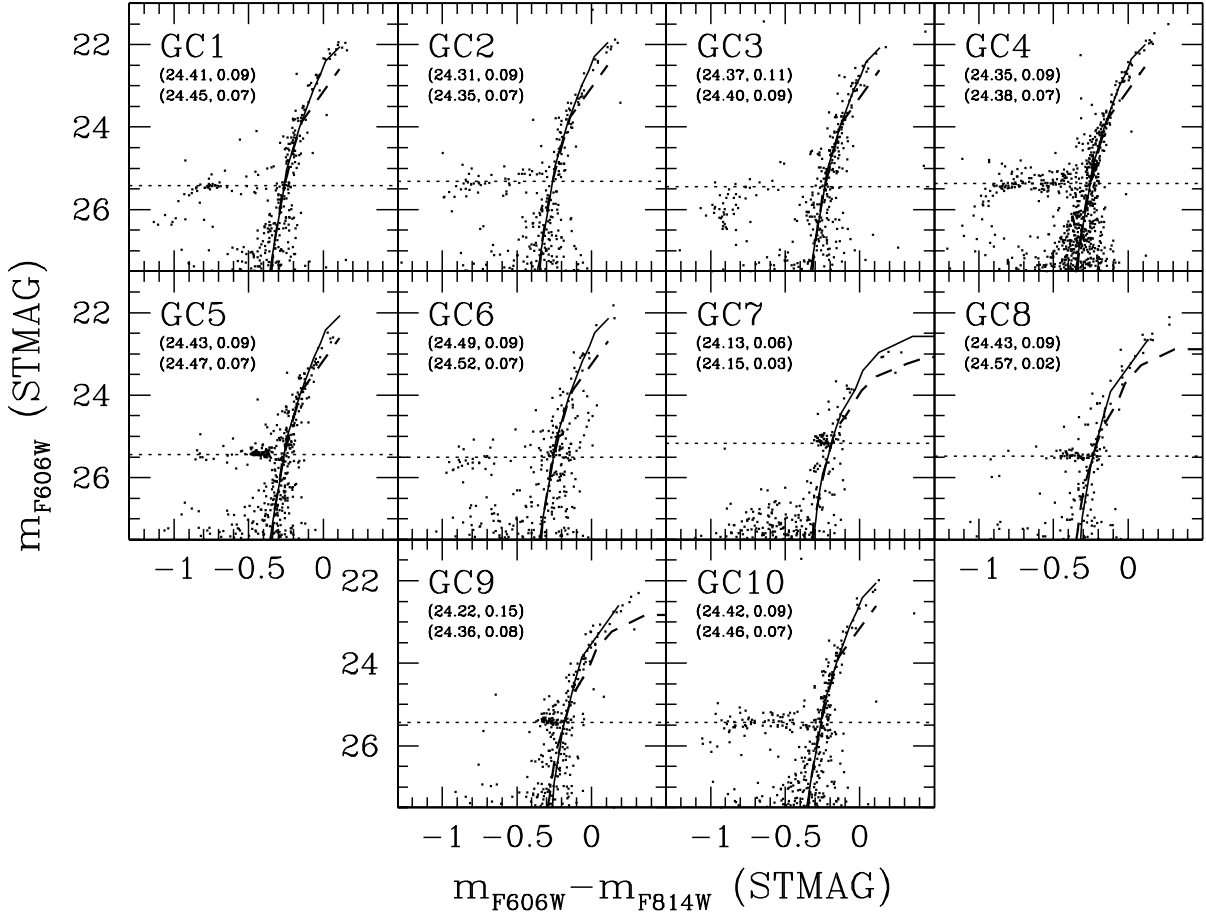}
\caption{Results of fitting the Galactic globular cluster fiducials from \citet{brown:05} to the observed CMDs. In all panels, the more metal-poor fiducial is marked with a solid line, and the more metal-rich fiducial with a dashed line. The numbers in the upper left of each panel indicate the best fitting $(m-M)_0$ and $E(B-V)$ for the two fiducials -- metal-poor above, and metal-rich below. The two marked fiducials for GC7 are 47 Tuc and NGC 5927, which have $[$Fe$/$H$]=-0.70$ and $-0.37$, respectively. For GC8 and GC9 the two fiducials are NGC 6752 ($[$Fe$/$H$]=-1.54$) and 47 Tuc, while for the remaining 
seven clusters the two fiducials are M92 ($[$Fe$/$H$]=-2.14$) and NGC 6752. Photometry has been transformed to STMAG to match the \citet{brown:05} measurements.\label{f:fids}}
\end{figure}

Fig. \ref{f:fids} shows the fiducial registration, while numerical results 
are listed in Table \ref{t:results}. The derived metallicities reflect those 
adopted for the reference clusters by \citet{brown:05}. Our typical measurement 
uncertainties are $\pm 0.15$ dex. 

The majority of our clusters are confirmed as metal-poor objects. Five have RGBs 
well traced by the M92 fiducial ($[$Fe$/$H$] = -2.14$), and two have RGBs bounded by 
M92 and NGC 6752 ($[$Fe$/$H$] = -1.54$). Two more have RGBs matched by the NGC 6752 
fiducial. The remaining cluster, GC7, is more metal-rich, with its RGB matched by 
47 Tuc ($[$Fe$/$H$] = -0.70$). Two clusters have previous metallicity measurements.
\citet{galleti:06} derived, photometrically, $[$Fe$/$H$] \sim -1.8$ for GC4. 
GC6 has a spectroscopic measurement by \citet{perrett:02}, who found 
$[$Fe$/$H$] = -2.07 \pm 0.11$. Both values are in reasonable agreement with our 
estimates.

We also compared our derived $E(B-V)$ with those from the maps of 
\citet[][SFD]{schlegel:98} (listed in Table \ref{t:results}). Agreement is adequate, 
although we note a tendency to derive larger $E(B-V)$ than SFD by a few
hundredths of a mag -- we noticed a similar effect
in Paper I. Adopting the SFD values forces our $(m-M)_0$ to be greater
by $\sim 0.025$ mag per $0.01$ mag difference in $E(B-V)$; however the fiducial 
registration is often noticeably inferior. The discrepancy
may be due to spatial reddening variations on smaller scales than resolved by the SFD
maps ($6\farcm1$), a systematic error introduced by slightly different ages for
the M31 and template clusters (as noted above), or to a systematic error in the 
colour excesses adopted by \citet{brown:05} for their reference clusters.

Most of our derived distance moduli lie close to the canonical value for M31: 
$(m-M)_0 = 24.47$ ($\sim 780$ kpc). However, our measurements for GC7 and GC9 
suggest these clusters may be closer by $\sim 115$ kpc and $\sim 85$ kpc, respectively. 
This would render GC7, in particular, an unusual object, given its $[$Fe$/$H$]$. 
In the Milky Way, there are two prominent globular clusters at unusually large radii for 
their $[$Fe$/$H$]$: Ter 7 and Pal 12. Both these clusters are only about $70 \%$ the
age of the oldest Galactic globulars. If GC7 is similarly young, then fitting the
\citet{brown:05} fiducials is inappropriate. We note that the CMD of GC7 shows
an overdensity of points around $m_{{\rm F606W}} \sim 27.5$ and
$m_{{\rm F606W}} - m_{{\rm F814W}} \sim 0.5$ which is not present
in the other cluster CMDs. The presence of the main-sequence turn-off in this
region, blurred by observational errors, could explain this bulge. If this is the case,
then GC7 may be as young as $\sim 4$ Gyr, based on CMDs for LMC and SMC clusters.
To check our measured $E(B-V)$ and $(m-M)_0$ for GC7, we transformed our photometry
to Johnson-Cousins $V$ and $I$ \citep{sirianni:05} and matched the fiducials of
\citet{sarajedini:97} and \citet{rosenberg:98} for Ter 7 and Pal 12, respectively.
These objects have comparable $[$Fe$/$H$]$ to GC7. The fiducial fits result in
$E(B-V) \sim 0.07$ and $(m-M)_0 \sim 24.3$, much more similar to the quantities
observed for our other M31 clusters. We note that both Ter 7 and Pal 12 are 
unambiguously associated with the accreted Sagittarius dSph. GC7 may represent a 
similar scenario in M31 -- verification of our measurements would therefore be valuable.

Inspection of Fig. \ref{f:cmds} and our measured $[$Fe$/$H$]$ reveals several
second parameter clusters. We defer numerical calculation of 
HB morphologies until a future paper when tests for completeness 
and photometric blends can be incorporated. Even so, the second parameter effect 
is clear for GC5, which has a red HB but equivalent $[$Fe$/$H$]$ to GC2, and EC1 
from Paper I. Similarly, GC8 and GC9 have 
$[$Fe$/$H$] \sim -1.54$; however they have much redder HBs than comparable 
Galactic globulars (e.g., NGC 1904, 6752, 7492). The HBs of these two objects 
resemble those for many of the M31 clusters with similar $[$Fe$/$H$]$ but smaller 
$R_p$ observed by \citet{rich:05}, and second parameter Galactic globulars 
such as Pal 14.

Finally, we calculated integrated cluster luminosities by means of
aperture photometry centred on the cluster centres. This technique ensures all 
light is counted, even in the unresolved cores of the most compact objects. For 
a given cluster, we first estimated the sky level using regions away from the
cluster, and produced a sky subtracted image. At large projected radii from M31,
we do not have to worry about a significantly spatially variable background.
We then masked any bright background galaxies or foreground stars in the vicinity
of the cluster. Next, we measured integrated luminosities using a variety of 
apertures. In combination with the cluster CMDs, plotting luminosity as a function
of aperture radius allowed us to check for field star contamination. For the most
remote clusters, where field contamination is extremely sparse, the integrated
luminosity quickly reached an asymptotic limit. For such objects a maximum radius
of $20\arcsec$ was more than sufficient. For objects set against more significant
fields, contamination manifested in the form of a slowly increasing integrated
luminosity with increasing aperture radius (as well as being visible in the CMD). 
For these objects, the cluster light always dominates for radii smaller than 
$10\arcsec$. Beyond this, we used our plots to estimate the radius at which field 
star contamination became non-negligible ($r_{{\rm max}}$) and integrated only to 
this limit. Three clusters have $r_{{\rm max}} = 10\arcsec$, while the remainder
have $16\arcsec \le r_{{\rm max}} \le 20\arcsec$. We used our summed
magnitudes in F606W and F814W to transform the F606W value to Johnson $V$ 
\citep{sirianni:05}, and then used the relevant $E(B-V)$ and 
$(m-M)_0$ to calculate $M_V$. We next re-sampled with smaller apertures to estimate
the half-light radius $r_h$, which 
we converted to parsecs using our $(m-M)_0$ values. All results are listed in
Table \ref{t:results}. Clusters with $r_{{\rm max}} \ga 16\arcsec$ have
reliable $M_V$ -- as noted above, the integrations are very close to asymptotic by 
such radii. For the three with $r_{{\rm max}} = 10\arcsec$, we used the profiles
of the clusters with negligible field contamination to observe that $M_V$ may be
under-estimated by $\sim 0.1$ mag. This level of uncertainty is acceptable for
present purposes -- it is less even than the uncertainties introduced by application
of the derived distance moduli.

With our new sample of remote M31 members, we are, for the first time,
in a position to compare the outer globular cluster system of this galaxy with
that of our own. Doing so reveals striking differences between the two.
Seven Galactic globular clusters have $R_{{\rm gc}} > 40$ kpc (Pyxis; Pal 3, 4,
and 14; AM-1; Eridanus; and NGC 2419). Of these, all except NGC 2419 have
$-1.8 \la [$Fe$/$H$] \la -1.3$, and are sub-luminous ($-6.0 \la M_V \la -4.7$) 
and very extended ($11 \la r_h \la 25$ pc). NGC 2419 is one of the most luminous
clusters in our Galaxy ($M_V \sim -9.6$), and lies in an unusual position on the luminosity 
vs. size plane. This has led to some suggestions it may be the remains of an accreted
galaxy \citep[e.g.,][]{vdb:04}. In our present sample, we have four M31 globulars at 
$R_p > 40$ kpc. In contrast with the majority of outer Milky Way clusters, these objects 
are very metal-poor ($-2.2 \la [$Fe$/$H$] \la -1.8$), compact ($4 \la r_h \la 7$ pc) and 
very bright ($-8.9 \la M_V \la -8.3$). Unlike NGC 2419, none of the clusters lie in an 
unusual position on the luminosity vs. size plane, suggesting that they are 'normal'
globular clusters. The new cluster of \citet{martin:06}, at $R_p \sim 116$ kpc is also 
compact and luminous ($M_V \sim -8.5$), while EC4 from Paper I, at $R_p \sim 60$ kpc, 
is a member of the population of luminous, extended M31 globular clusters which
has no Galactic analogue. 

It is perhaps not surprising that we have not observed any very extended, 
sub-luminous globular clusters in the remote M31 halo -- such objects may
well lie below our present survey detection limit. Even so, the above comparison 
shows that {\it M31 clearly possesses an extended system of metal-poor, compact, 
and very luminous globular clusters which is not seen in the Milky Way}. The one
luminous, metal-poor outer Milky Way cluster -- NGC 2419 -- is quite unlike the M31 
clusters we have observed (as described above). Furthermore, 
the outermost M31 globular clusters studied here are considerably more metal-poor
than the pressure-supported field halo population at the same radii 
\citep{chapman:06,kalirai:06}, again in contrast to the Milky Way. These
disparities may well offer important clues to differences in the early formation 
and evolution of the two galaxies or in their subsequent accretion histories, and as
such, it is vital that additional M31 members are sought. We remark that the 
present sample has been obtained from a survey of $\approx 100$ square degrees, 
roughly 25\% of the area contained within a projected radius of 150~kpc of M31. 
A significant number of new clusters may therefore await detection, suggesting that,
unlike in the Milky Way, there may be a relatively large population of luminous, compact 
globular clusters in the outer M31 halo. Such objects, in addition to their value as 
globular clusters, are extremely useful as dynamical probes of the mass distribution in M31 -- 
work which has previously relied on clusters with $R_p \la 25$ kpc and a handful 
of dwarf galaxies. We anticipate that our new sample will soon allow 
improved mass constraints to be determined.

\acknowledgments
ADM and AMNF are supported by a Marie Curie Excellence Grant from the European 
Commission under contract MCEXT-CT-2005-025869. NRT acknowledges financial
support via a PPARC Senior Research Fellowship.



{\it Facilities:} \facility{HST (ACS)}.


\clearpage

\begin{deluxetable}{lcccccccccc}
\tabletypesize{\footnotesize}
\rotate
\tablecaption{Observed properties of ten globular clusters in the halo of M31\label{t:results}}
\tablewidth{0pt}
\tablehead{
\colhead{Identifier\tablenotemark{a}} & \colhead{RA} & \colhead{Dec} & \colhead{$R_p$} & \colhead{$(m-M)_0$} & \colhead{$E(B-V)$} & \colhead{$E(B-V)$} & \colhead{$[$Fe$/$H$]$} & \colhead{$r_h$} & \colhead{$M_V$} & \colhead{$r_{{\rm max}}$} \\
 & \colhead{(J2000.0)} & \colhead{(J2000.0)} & \colhead{(kpc)} & & \colhead{(meas.)} & \colhead{(lit.)} & & \colhead{(pc)} & & \colhead{($\arcsec$)}}
\startdata
GC1 & $00^{{\rm h}} 26^{{\rm m}} 47\fs8$ & $+39\degr 44\arcmin 45\farcs5$ & $46.4$ & $24.41 \pm 0.14$ & $0.09 \pm 0.01$ & $0.07$ & $-2.14$ & $3.8$ & $-8.7$ & $20$ \\
GC2 & $00^{{\rm h}} 29^{{\rm m}} 44\fs9$ & $+41\degr 13\arcmin 09\farcs8$ & $33.4$ & $24.32 \pm 0.14$ & $0.08 \pm 0.01$ & $0.07$ & $-1.94$ & $5.2$ & $-7.7$ & $16$ \\
GC3 & $00^{{\rm h}} 30^{{\rm m}} 27\fs3$ & $+41\degr 36\arcmin 20\farcs4$ & $31.8$ & $24.37 \pm 0.15$ & $0.11 \pm 0.01$ & $0.07$ & $-2.14$ & $9.9$ & $-8.5$ & $16$ \\
GC4 & $00^{{\rm h}} 31^{{\rm m}} 09\fs9$ & $+37\degr 53\arcmin 59\farcs7$ & $55.2$ & $24.35 \pm 0.14$ & $0.09 \pm 0.01$ & $0.06$ & $-2.14$ & $6.8$ & $-8.9$ & $20$ \\
GC5 & $00^{{\rm h}} 35^{{\rm m}} 59\fs7$ & $+35\degr 41\arcmin 03\farcs6$ & $78.5$ & $24.45 \pm 0.15$ & $0.08 \pm 0.01$ & $0.07$ & $-1.84$ & $6.3$ & $-8.8$ & $18$ \\
GC6 & $00^{{\rm h}} 38^{{\rm m}} 00\fs3$ & $+40\degr 43\arcmin 56\farcs1$ & $14.0$ & $24.49 \pm 0.15$ & $0.09 \pm 0.01$ & $0.07$ & $-2.14$ & $3.9$ & $-8.4$ & $10$ \\
GC7 & $00^{{\rm h}} 38^{{\rm m}} 49\fs4$ & $+42\degr 22\arcmin 48\farcs0$ & $18.2$ & $24.13 \pm 0.13$ & $0.06 \pm 0.01$ & $0.06$ & $-0.70$ & $7.5$ & $-6.2$ & $10$ \\
GC8 & $00^{{\rm h}} 54^{{\rm m}} 25\fs0$ & $+39\degr 42\arcmin 55\farcs5$ & $37.1$ & $24.43 \pm 0.14$ & $0.09 \pm 0.01$ & $0.05$ & $-1.54$ & $3.2$ & $-8.0$ & $10$ \\
GC9 & $00^{{\rm h}} 55^{{\rm m}} 44\fs0$ & $+42\degr 46\arcmin 16\farcs1$ & $38.9$ & $24.22 \pm 0.14$ & $0.15 \pm 0.01$ & $0.10$ & $-1.54$ & $9.8$ & $-7.2$ & $16$ \\
GC10 & $01^{{\rm h}} 07^{{\rm m}} 26\fs4$ & $+35\degr 46\arcmin 49\farcs7$ & $99.9$ & $24.42 \pm 0.14$ & $0.09 \pm 0.01$ & $0.05$ & $-2.14$ & $4.3$ & $-8.3$ & $20$
\enddata
\tablenotetext{a}{Two clusters are previously labelled in the literature: GC4 is B514 in \citet{galleti:05,galleti:06}; GC6 is 298-021 in \citet{barmby:00}.}
\end{deluxetable}

\clearpage

\clearpage

\clearpage

\end{document}